\documentclass[aps,pra,eqsecnum,superscriptaddress,groupedaddress,nofootinbib]{revtex4-2}

\usepackage{amssymb,amsmath}
\usepackage{subdepth}
\usepackage{graphicx}
\usepackage{sansmath}
\usepackage{bm}
\usepackage{soul}
\usepackage{framed}

\usepackage[breaklinks=true]{hyperref}
\hypersetup{
	colorlinks   = true, %Colours links instead of ugly boxes
	urlcolor     = blue, %Colour for external hyperlinks
	linkcolor    = red, %Colour of internal links
	citecolor   = green %Colour of citations
}

\DeclareMathOperator{\Tr}{Tr}

\newcommand{\ket}[1]{\vert#1\rangle}
\newcommand{\bra}[1]{\langle#1\vert}

\newcommand{\proj}[1]{| #1\rangle\!\langle #1 |}

\newcommand{\inv}{{\,\text{-}\hspace{-1pt}1}}

\newcommand{\Z}{\mathcal{Z}}

\newcommand{\J}{\mathcal{J}}

\newcommand{\UU}{\mathcal{U}}

\newcommand{\E}{\mathcal{E}}%For the love of Cartan

\renewcommand{\O}{\mathcal{O}}

\newcommand{\ZZ}{\mathbb Z}

\newcommand{\R}{\mathbb R}
\newcommand{\C}{\mathbb C}

\usepackage{color}

\begin{document}
	
	\title{The photodetector, the heterodyne instrument,\\ and the principle of instrument autonomy}
	
	\author{Christopher S. Jackson}
	\email{omgphysics@gmail.com}
	\affiliation{Quantum Algorithms and Applications Collaboratory, Sandia National Laboratories, Livermore, CA 94550, USA}

	\date{\today}
	
\begin{abstract}
	Measuring instruments, especially ones that observe continually over time, have a reality to them that is independent of the states that stimulate their senses.
	This is the Principle of Instrument Autonomy.
	Although the mathematical concept of an instrument implicitly embodies this principle, the conventional analysis of continual observation has become overwhelmingly focused on state evolution rather than on descriptions of instruments themselves.
	Because of this, it can be hard to appreciate that an instrument that observes for a finite amount of time has an evolution of its own, a stochastic evolution that precedes the Born rule and Schr\"odinger equation of the measured system.
	In this article, the two most established of the continually observing instruments, the Srinivas-Davies photodetector and the Goetsch-Graham-Wiseman heterodyne detector, are reviewed with an emphasis on the autonomous instrument evolution they define, made explicit by application of the recently introduced Kraus-operator distribution function.
	It is then pointed out how the heterodyne instrument evolution is a complete alternative to the original idea of energy quantization, where the usual ideas of \emph{temperature} and \emph{energy} of a \emph{state} are replaced by the \emph{time} and \emph{positivity} of the \emph{instrument}.
	%Further, it is explained how this concept of instrument evolution is related to the concepts of renormalization group in field theory and suggested that the time of a continual observation offers another scheme for regularization.
\end{abstract}
	
	%\pacs{}
	\maketitle

\section{Introduction: Observables, Observation, and Observers}

From a quantum theoretical point-of-view, observations are equal to changes of state or so-called \emph{operations}.
An \emph{instrument} is by definition a set of operations consistent with statistics and a \emph{measurement} or \emph{observation} is the selection of one of those operations.
There are two extremes for where to place an instrument with respect to the observer.
In open systems theory, the instrument is considered to be hidden from the observer.
In measurement theory, the instrument is considered to be all that the observer can sense, and this will be the current setting.
A very important assumption about measurement is that the process by which an observation is selected is required to be fundamentally random | that is, God plays dice after all~\cite{wheeler2014quantum}.
This requirement for randomness in order to qualify as a measurement represents a fundamental principle about the purpose of measurement:
the event is an opportunity for something new to be learned and is therefore out of the observer's control~\cite{fuchs2014quantum}.

The technical apparatus of quantum measurement theory took quite a while to develop fully | almost sixty years between the Heisenberg Uncertainty Principle, published 1927~\cite{wheeler2014quantum}, and Kraus's lectures, published 1983~\cite{kraus1983states}.
Today, about 40 years after its essential completion, the dissemination and application of this advanced theory is still a work in progress.
If I were to summarize what this advanced theory accomplished, I would say it gave the capacity to discuss the topic of continual observation | that is, the relationship between measuring processes and time.
On the topic of continual observation, there are two models that are the most developed and established, grounding the whole discussion:
The Srinivas-Davies photodetector~\cite{srinivas1981photon} and the Goetsch-Graham-Wiseman heterodyne detector~\cite{wiseman1994quantum,Goetsch1994a,wiseman1996measurement}.

The Srinivas-Davies model of photodetection describes the continual observation of quantum energy or ``photon number", an actual measuring process in time that upon completion, gives the traditional idea of measurement implied by the Hermitian number operator, $a^\dag a$.
Whereas the traditional idea simply appreciates what it means for the eigenstates of the number observable to resolve the identity of the Hilbert space,
\begin{equation}
	1 = \sum_{n=0}^\infty \proj{n},
\end{equation}
the Srinivas-Davies model essentially expands on these eigenvalues as a function of time, where they appear as the total counts of a Poisson process,
\begin{equation}
	n = \int_0^\infty dp_t,
\end{equation}
where $dp_t$ is the observed Poisson process.

The Goetsch-Graham-Wiseman model of heterodyne describes the continual observation of both amplitude quadratures simultanously, an actual measuring process in time that upon completion, gives the measurement implied by the ``overcomplete basis"~\cite{klauder1960action} of coherent states~\cite{glauber1963coherent}.
This kind of observation is different from the traditional idea of the Hermitian observable because it is a simultaneous observation of the canonical Heisenberg observables, ``$Q$ and $P$'', which famously don't commute.
Of course, this sort of simultaneous observation does not contradict the Heisenberg uncertainty principle, precisely because it is a continual observation, weak for small times.
Previous to the Goetsch-Graham-Wiseman model, this idea of measurement was only suggested by the so-called ``overcomplete basis'' of coherent states,
\begin{equation}\label{CSPOVM}
	1 = \int_\C \frac{d^2\alpha}{\pi}\proj{\alpha}.
\end{equation}
With the Goetsch-Graham-Wiseman model, it becomes clear how such a ``spectrum'' of complex amplitudes can be observed as they appear in the form of an integrated complex Wiener processes,
\begin{equation}
	\alpha = \int_0^\infty \sqrt{\kappa_o} dw_t e^{-\frac12\kappa_o t},
\end{equation}
where $\kappa_o$ is the observation rate and $dw_t$ is the observed complex Wiener process, with real and imaginary parts corresponding to the observables ``$Q$ and $P$'' in the interaction picture.\\

With regard to the broader topic of continual observation, I would now like to raise the main concern of this article: it appears that measurement theory today has become overwhelmed by the discussion of state evolution via equations such as the nonlinear Schrodinger equation, the Lindblad master equation, the stochastic master equation, or linear trajectories.
Although it is true that quantum observations are ultimately equal to changes of state, this does not mean that state evolution is the subject of measurement theory.
It must not be forgotten, and I beg readers to consider this, that instruments (that is, literal measuring devices) have a reality and therefore potential life of their own, independent of the states that stimulate their senses.
Specifically, this means that any instrument which continually observes over time defines an evolution process that is actually more fundamental than state evolution.
That this type of non-state  evolution exists is exactly what I hope readers will take the time to consider.

If you are asking yourself ``What on Earth is evolving if not the state?'', consider what it means for a measuring instrument to be a set of possible operations.
During a continual observation, there will always be multiple possible measurement records which become assigned to the same operation, and this multitude of possible records per operation defines a distribution.
By virtue of their composition, the operations of an instrument for an infinitesimal amount of time defines a \emph{domain} of operations at any finite amount of time.
That is, continual observations define instruments with the so-called Markov or group property.
In four turns, this domain is (first) a generally multi-dimensional semigroup which (second) has a Haar measure, and it is with respect to this measure that (third) the distribution of operations and (fourth) its autonomous evolution are defined.

This idea of an evolving Haar-based distribution of operations was introduced recently for the purpose of showing how  the ``measurement in the spin-coherent-state basis'' can be accomplished by means of an isotropic detection scheme~\cite{Jackson2021perform}.
There it was called the \emph{Kraus-operator distribution function} because the operations of the isotropic instrument are Kraus-rank one.
So too in this article, the Kraus-operator distribution function and its evolution will be explained for the Srinivas-Davies and Goetsch-Graham-Wiseman instruments.\\

Before we begin, I would like to emphasize two more points about Markovian measuring processes:
Although it is widely known that the Markovianity of continual observations corresponds to a semigroup of operations,
it appears to be just as widely mistaken that this semigroup can only refer to the single parameter of time, conjugate to the infinitesimal generator of the total operation~\cite{lindblad1976generators}.
So my first point is that the selective operations (that is, the elements of the instrument) themselves also form a semigroup, adding further degrees of freedom to the single parameter of time | e.g. the Srinivas-Davies instrument includes a semigroup of natural numbers and the Goetsch-Graham-Wiseman instrument includes a group of complex amplitudes.
The second point is that although there has been a great deal of interest in adaptive measurement and feedback control~\cite{wiseman1994quantum, doherty2000quantum}, nonadaptive measurement is not to be mistaken as an understood theory.
Indeed, nonadaptive measurement is not complete because there hasn't been sufficient work on the various noncommutative structures that must generally be considered for the continual observation of multiple simultaneous observables.

\section{Advanced Measurement Theory}

We will only need to consider operations of Kraus-rank one, so an \emph{operation} can be defined as a superoperator of the form
\begin{equation}
	\O(K) = K \!\odot\! K^\dag,
\end{equation}
where $K$ is an operator on the system Hilbert space and the ``odot'' is a tensor product which implies that the operation acting on a state should be contracted as
\begin{equation}
	A \!\odot\! B \,(\rho) \equiv A\rho B,
\end{equation}
where $\rho$ is the state's density operator.
An \emph{instrument} is then defined as a set of operations,
\begin{equation}
	\mathcal{I} =\Big \{\O(K_x)\Big\},
\end{equation}
which together satisfy the \emph{completeness} relation,
\begin{equation}
	\sum_x K_x ^\dag K_x = 1,
\end{equation}
and the operators $K_x$ are then called \emph{Kraus operators}.
Upon the measurement of a state, an element $x$ is selected.
The probability of this selection to register is given by the \emph{(advanced) Born rule},
\begin{equation}
	P(x|\rho) = \Tr(K_x^\dag K_x \rho),
\end{equation}
and the state is understood to transition to $K_x \rho K_x^\dag$, renormalized.
By \emph{the second Kraus representation theorem}, an instrument can always be expanded in terms of the two more traditional objects of quantum theory, the unitary interaction and the measurement of a Hermitian observable.
That is, each Kraus operator can be considered to be of the form
\begin{equation}
	K_x = \bra{x}\UU\ket{0},
\end{equation}
where a meter is initiated in the pointer state $\ket{0}$  (pure because we are only considering Kraus-rank-one operations), is coupled to the system via a system-meter unitary $\UU$, and is terminated in a register of the eigenvalue $x$ with linear functional $\bra{x}$.

For every instrument, there are two subsequent objects that are also fundamental to consider.
On the one hand, there is the set of effects
\begin{equation}\label{POVM}
	\E = \Big\{K_x ^\dag K_x\Big\},
\end{equation}
which is all that is needed if the observer does not need to continue their observations.
Sets of operators of the form \ref{POVM} have the distinct property that they consist of positive operators which sum to the identity and are thus also called (rather uncreatively) Positive-Operator-Valued Measures or \emph{POVM}s.
On the other hand, there is the total operation
\begin{equation}\label{CPTP}
	\Z = \sum_x \O(K_x),
\end{equation}
which is all that is important if the register is hidden from the observer.
Superoperators of the form~\ref{CPTP} have the distinct property that for any appended Hilbert space with identity superoperator $\mathcal{I}d$, $\Z \otimes \mathcal{I}d$ maps normalized positive operators to normalized positive operators and are thus also called Completely Positive and Trace Preserving or \emph{CPTP} map.
By \emph{the first Kraus representation theorem}, a CPTP map can always be expanded or ``unraveled'' in terms of an instrument as in equation~\ref{CPTP}.\\

To finish this run through advanced measurement theory, I'd like to repeat my concern about the culture of its practice.
Many significant developments have happened in the understanding of CPTP superoperators~\cite{choi1975completely, lindblad1976generators, davies1976quantum,  kraus1983states, nielsen2002quantum, kitaev2002classical,bengtsson2006geometry,wiseman2009quantum}.
However, these are not developments in the direct understanding of instruments themselves.
It appears to me that a direct study of instruments is often avoided, and I believe this is because it is hard to step away from the two very influential ideas embedded in the mathematical theory of quantum mechanics as it was originally designed by von Neumann~\cite{vonneumann1932mathematical}.
Specifically, these are the idea of linear basis invariance, which the Hilbert space was introduced to celebrate, and the idea of measuring processes, which was introduced to interpret thermal equilibrium.
So to finish this section, I'd like to remind the reader:
This article is not about open systems;
it is about measuring devices, particularly photodetectors and heterodyne instruments.

%The total unconditional process is the usual
%\begin{align}
%	\Z_{dt} &= K_0\odot K_0^\dag + K_1\odot K_1^\dag\\
%	&= \exp\left(-i\omega_o\ad_{a^\dag a} dt +\D(a) dt  \right)
%\end{align}
%where defined is the dissipator
%\begin{equation}
%	\D(a) = \kappa_o \left(a\odot a^\dag - \frac12\left(a^\dag a \odot 1 + 1 \odot a^\dag a\right)\right)
%\end{equation}
%in which case the standard Lindblad operator is $\sqrt{\kappa_o}a$.
%Although the exponential form of the unconditional process is mentioned here, it will not be made much use of and it is only mentioned to give something to look at that may be more familiar to many readers.

\section{The Srinivas-Davies Photodetector}

The Srinivas-Davies photodetector is a model that describes the changes in state of an electromagnetic field mode, called the leaky cavity, when such an instrument is in vacuum, looking continually in time for photons to ``drop'' from the leaky cavity, with the interaction between the cavity and the instrument vacuum being that assumed by the rotating-wave approximation.
Much like if one listens for raindrops, the possible observations of the photodetector are Poisson processes, $dp_t$, which can be expressed in terms of the number of drops and the time of each drop,
\begin{equation}
	dp_t = dt \sum_{k=1}^n \delta(t-T_k),
\end{equation}
where $\delta(t)$ is the Dirac delta distribution.
When a photon drops, the state of the cavity in the interaction picture is changed by the Kraus operator
\begin{equation}
	K_1 = a \sqrt{\kappa_o dt},
\end{equation}
where $\kappa_o$ is the infinitesimal observation rate, $dt$ is the temporal resolution, and $a$ is the standard lowering operator.
When a photon doesn't drop, the state of the cavity is still changed, but by the Kraus operator
\begin{equation}
	K_0 = e^{-a^\dag a\, \frac12 \kappa_o dt}.
\end{equation}
This detail about the effect of observing no photon is required by probability and is precisely the point that Srinivas and Davies~\cite{srinivas1981photon} appear to be the first to make correctly.
The most important aspect of these zero-photon Kraus operators is the measurement interpretation they give to the identity
\begin{equation}\label{renormalizationSD}
	\boxed{
		\vphantom{\Bigg(}
		\hspace{15pt}
		a\; e^{-a^\dag a\, \frac12 \kappa_o  T}
		= e^{-a^\dag a\, \frac12 \kappa_o  T}  a \,Z_\alpha^{-1/2},
		\hspace{10pt}
		\vphantom{\Bigg)}
	}
\end{equation}
where $Z_\alpha^{-1/2}=e^{-\frac12\kappa_o T}$ is a renormalization factor.
Stating the measurement interpretation in verse:
\begin{verse}\centering
	\fbox{\emph{The field amplitude measured by the photodetector is effectively renormalized over time.}}
\end{verse}
That is, a photon registering at time $t=T$ is equal to a photon registering at time $t=0$ with the amplitude scaled by the renormalization factor, $Z_\alpha^{-1/2}=e^{-\frac12\kappa_o T}$.

As the photodetector observes for a finite amount of time $T$, the Kraus operators are composed in order of time,
\begin{equation}
	K[dp_{[0,T)}]= \mathrm{T}\prod_{t=0}^{T/dt-1}K_{dp_t}
	\equiv K_{p_{T-dt}} \cdots K_{p_{dt}} K_{p_{0}},
\end{equation}
where $\mathrm{T}$ denotes time order.
With liberal application of equation~\ref{renormalizationSD}, each operation can be put into a standard order,
\begin{align}
	\O\Big(K\left[dp_{[0,T)}\right]\Big)
	&= \O\bigg(e^{- a^\dag a \frac12\kappa_o (T-T_n)} \left(\sqrt{\kappa_o dt}a\right) \cdots \left(\sqrt{\kappa_o dt}a\right) e^{-a^\dag a \frac12 \kappa_o (T_2-T_1)}\left(\sqrt{\kappa_o dt}a\right) e^{- a^\dag a \frac12\kappa_o T_1}\bigg)\\
	&= (\kappa_o dt)^ne^{-\kappa_o (T_n+\cdots +T_1)}\O\left(e^{- a^\dag a \frac12\kappa_o T}a^n\right),
\end{align}
and the instrument can be partitioned into equivalence classes enumerated by the total number of photons,
\begin{align}
	\Z_T
	&= \sum_{dp_{[0,T)}}\O\Big(K\left[dp_{[0,T)}\right]\Big)\\
	&=\sum_{n=0}^\infty\sum_{\;T>T_n > \cdots > T_1\ge0}
	(\kappa_o dt)^n e^{-\kappa_o (T_n+\cdots +T_1)}
	\O\left(e^{-a^\dag a\, \frac12 \kappa_o  T}a^n\right)\\
	&\equiv \sum_{n=0}^\infty \tilde{D}_T(n)
	\O\left(e^{-a^\dag a\, \frac12 \kappa_o  T}a^n\right),
\end{align}
where defined is the unnormalized \emph{Kraus-operator distribution function},
\begin{align}
	\tilde{D}_T(n) &\equiv \sum_{\;T>T_n > \ldots > T_1\ge0}
	(\kappa_o dt)^n e^{- \kappa_o (T_n+\cdots +T_1)}\\
	&= \int_0^T\kappa_o dT_n\cdots \int_0^{T_2}\kappa_o dT_1 e^{- \kappa_o (T_n+\cdots +T_1)}\\
	&= \int_{e^{-\kappa_o T}}^1d(e^{-\kappa_o T_n})\cdots \int_{e^{-\kappa_o T_2}}^1 d(e^{-\kappa_o T_1})\\
	&= \int_0^{1-e^{-\kappa_o T}} dx_n\cdots \int_0^{x_2} dx_1\\
	&= \frac{1}{n!}\left(1-e^{-\kappa_o T}\right)^n,
\end{align}
which can be recognized as an unnormalized Poisson distribution, as had been expected, except that the effective mean,
\begin{equation}
	\lambda(T) = 1-e^{-\kappa_o T}=\kappa_o\int_0^T dt\,e^{-\kappa_o t},
\end{equation}
is not linear in time precisely because of the required amplitude renormalization.

\subsection*{Summary}

The Srinivas-Davies photodetector can thus be summarized as a partition of operations,
\begin{align}
	\Z_T = \sum_{n=0}^\infty D_T(n)\,\O\Big(K_T(n)\Big),
\end{align}
with Kraus operators,
\begin{equation}
	K_T(n) = e^{\frac12\lambda(T)}e^{-a^\dag a\, \frac12 \kappa_o  T}a^n,
\end{equation}
conjugate to a normalized \emph{Kraus-operator distribution function},
\begin{align}
D_T(n)=\tilde{D}_T(n)e^{-\lambda(T)}=e^{-\lambda(T)}\frac{\lambda(T)^n}{n!},
\end{align}
which evolves according to the equation
\begin{equation}\label{Poisson}
	\boxed{
		\vphantom{\Bigg(}
		\hspace{15pt}
		\frac{d}{dt}D_t(n) = \kappa(t)\Big(D_t(n-1)-D_t(n)\Big),
		\hspace{10pt}
		\vphantom{\Bigg)}
	}
\end{equation}
where $\kappa(t)=\dot\lambda(t)=\kappa_o e^{-\kappa_o t}$ is an effective observation rate that modifies the Poisson distribution of Kraus operators and the initial condition is $D_0(n)=\delta_{n,0}$.
This modification to the Poisson process is so important that it, too, is worth stating in verse:
\begin{verse}\centering
	\fbox{\emph{The coupling of the photodetector to the cavity is effectively screened over time.}}
\end{verse}
That is, the effective coupling at $t=T$ is equal to the coupling at $t=0$
rescaled by the renormalization factor, $Z_\alpha^{-1}=e^{-\kappa_o T}$.

This modification to the Poisson distribution of Kraus operators was the main result of~\cite{srinivas1981photon}, although the Kraus-operator distribution function is not stressed there like it is here.
It is very important to acknowledge the existence of this Kraus-operator distribution function and its evolution, as it describes essential features of the photodetector, features that are independent of whatever state may stimulate it.
In other words, the Kraus-operator distribution function is an \emph{autonomous} description of the photodetector.
Of course, the Kraus-operator distribution function is not to be confused with the distribution of actual samples a state would cause, but rather is a factor that precedes the distribution of samples predicted by the Born rule,
\begin{equation}
	P_T(n|\rho) = D_T(n)\Tr\!\left(K_T(n)^{\dag} K_T(n)\,\rho\right).
\end{equation}
In this context, the Kraus-operator distribution function is an example of what has been called a \emph{method C ostensible distribution} by Wiseman~\cite{wiseman1996measurement} and this will be discussed more in the Conclusion.
Further, it should be noticed that the POVM elements are still quite similar to the multimode Mandel formula,
\begin{align}
D_T(n)K_T(n)^{\dag} K_T(n)
&=\frac{\lambda(T)^n}{n!}a^{\dagger n}e^{- a^\dagger\! a\,\kappa_o T}a^n\label{firstformSD}\\
&={\bf :}\frac{\lambda(T)^n}{n!}(a^\dagger a)^n e^{-a^\dagger\! a\,\lambda(T)}{\bf :}\,,\label{secondformSD}
\end{align}
where the bracketing colons denote normal ordering and equation \ref{secondformSD} makes obvious the POVM satisfies the completeness relation.
Finally, it is easy to see from equation \ref{firstformSD} that the POVM elements are
\begin{equation}
	D_T(n)K_T(n)^{\dag} K_T(n) = \proj{n} +O\!\left(n\,e^{-\kappa_oT}\right),
\end{equation}
so that the time of the Srinivas-Davies photodetector can be seen as a continuous bridge from the traditional idea of measuring the number observable back to the equally fundamental idea of not measuring.

\section{The Goetsch-Graham-Wiseman Heterodyne Instrument}\label{hetero}

The Goetsch-Graham-Wiseman heterodyne instrument is a model that describes the changes of state in a cavity when such an instrument is in vacuum, mixed with a (classical) local oscillator, listening continually in time to the amplitude and phase of the cavity, with the interaction between the cavity and the instrument vacuum being that assumed by the rotating-wave approximation.
Much like listening to the static of a radio, the possible observations of the heterodyne detector are complex Wiener processes, $dw_t$.
At every moment in time, a complex Wiener increment $dw$ is registered, and the state of the cavity is changed by the corresponding Kraus operator
\begin{equation}
	K(dw) = \sqrt{d\mu(dw)}L(dw),
\end{equation}
where defined are the complex Weiner measure of the increment
\begin{equation}
	d\mu(dw) = \frac{d^2(dw)}{\pi dt}e^{dw^*dw/dt},
\end{equation}
the conjugate operators
\begin{equation}
	L(dw) = e^{-\frac12 a^{\!\dag} \!a\, \kappa_o dt+a \sqrt{\kappa_o} dw^*},
\end{equation}
the infinitesimal observation rate $\kappa_o$, and the temporal resolution $dt$.
The point that these Kraus operators model heterodyne, a technology found in telecommunication that is actually older than quantum mechanics, appears in the literature to have been first mentioned by Wiseman~\cite{wiseman1994quantum}, thoroughly brought forward by Goetsch and Graham~\cite{Goetsch1994a}, and then further integrated and developed by Wiseman~\cite{wiseman1996measurement}.
Similar to the photodetector, a specific identity becomes the most important for continual observation,
\begin{equation}\label{renormalizationW}
	\boxed{
		\vphantom{\Bigg(}
		\hspace{15pt}
		e^{a\, dw^*}e^{-\frac12 a^{\!\dag} \!a\, \kappa_o T}
		= e^{-\frac12 a^{\!\dag} \!a\, \kappa_o T}e^{a\, Z_\alpha^{-1/2} dw^*},
		\hspace{10pt}
		\vphantom{\Bigg)}
	}
\end{equation}
where again $Z_\alpha^{-1/2}=e^{-\frac12\kappa_o T}$ is a renormalization factor.
Clearly, the exact same feature occurs except that the complex amplitude operator appears as an infinitesimal generator instead of as the base of an integer power.
So the identity can be interpreted by exactly the same verse as before (modulo the change in instrument):
\begin{verse}\centering
	\fbox{\emph{The field amplitude measured by the heterodyne instrument is effectively renormalized over time.}}
\end{verse}
That is, an increment registering at time $t=T$ is equal to an increment registering at time $t=0$ with the increment scaled by the renormalization factor $Z_\alpha^{-1/2}=e^{-\frac12\kappa_o T}$.

As the heterodyne instrument continually observes for a finite amount of time $T$,
the Kraus operators are composed in order of time,
\begin{align}
	K[dw_{[0,T)}] = \mathrm{T}\prod_{t=0}^{T/dt-1}K(dw_t)
	= \sqrt{\mathcal{D}\mu \left[dw_{[0,T)}\right]}\;\mathrm{T} \exp\left(\int_0^T -\frac12 a^{\!\dag} \!a\, \kappa_o dt+a \sqrt{\kappa_o} dw_t^*\right),
\end{align}
where defined is the Wiener path measure,
\begin{equation}
	\mathcal{D}\mu\!\left[dw_{[0,T)}\right] = \prod_{t=0}^{T/dt-1}d\mu(dw_t).
\end{equation}
Applying equation~\ref{renormalizationW} allows once again each operation to be put into a standard order
\begin{align}
	\O\left(K[dw_{[0,T)}]\right) = \mathcal{D}\mu\! \left[dw_{[0,T)}\right]\O\!\left(e^{-\frac12 a^{\!\dag} \!a\, \kappa_o T}e^{a\,\mu[dw_{[0,T)}]^* }\right),
\end{align}
where defined is the fundamental linear functional of the heterodyne process
\begin{equation}\label{heterofunc}
	\mu\!\left[dw_{[0,T)}\right]=\int_0^T \!\!\sqrt{\kappa_o} dw_te^{-\frac12\kappa_o t}.
\end{equation}
The instrument partitions into equivalence classes enumerated by the value of this functional,
\begin{align}
	\Z_T
	&= \sum_{dw_{[0,T)}}\O\left(K[dw_{[0,T)}]\right)\\
	&=\int\mathcal{D}\mu\! \left[dw_{[0,T)}\right]\O\!\left(e^{-\frac12 a^{\!\dag} \!a\, \kappa_o T}e^{a \,\mu[dw_{[0,T)}]^*  }\right)\\
	&=\int\mathcal{D}\mu\! \left[dw_{[0,T)}\right]\O\!\left(e^{-\frac12 a^{\!\dag} \!a\, \kappa_o T}e^{a \,\mu[dw_{[0,T)}]^*  }\right)\int \frac{d^2\zeta}{\pi}\,\pi\delta^2\!\Big(\zeta-\,\mu\left[dw_{[0,T)}\right] \Big)\\
	&=\int\frac{d^2\zeta}{\pi}D_T(\zeta)\;\O\!\left(e^{-\frac12 a^{\!\dag} \!a\, \kappa_o T}e^{a \zeta^*}\right),
\end{align}
and this partition defines an (already normalized) \emph{Kraus-operator distribution function},
\begin{align}
	D_T(\zeta)
	&=\int\mathcal{D}\mu\! \left[dw_{[0,T)}\right]\,\pi\delta^2\!\Big(\zeta-\,\mu\left[dw_{[0,T)}\right] \Big)\\
	&=\frac{1}{\Sigma(T)}e^{-\zeta^*\zeta/\Sigma(T)}\label{Gaussian},
\end{align}
which is obviously Gaussian with complex-isotropic covariance, as is to be expected, except that the effective covariance,
\begin{align}
	\Sigma(T) &= \int\mathcal{D}\mu\! \left[dw_{[0,T)}\right]\,\Big|\mu\!\left[dw_{[0,T)}\right]\!\Big|^2\\
	&= \kappa_o\int_0^T\int_0^T \left(\int\mathcal{D}\mu\! \left[dw_{[0,T)}\right]\, dw_s^*dw_t\right)e^{-\frac12\kappa_o (s+t)}\\
	&= \int_0^T \!\!\kappa_odt\,e^{-\kappa_o t}\\
	&= 1-e^{-\kappa_o T},
\end{align}
is not linear in time (as it would be for Brownian diffusion) precisely because of the amplitude renormalization.

\subsection*{Summary}

The Goetsch-Graham-Wiseman heterodyne instrument can be summarized as a partition of operations,
\begin{equation}
	\Z_T =\int\frac{d^2\zeta}{\pi}D_T(\zeta)\;\O\Big(K_T(\zeta)\Big),
\end{equation}
with Kraus operators,
\begin{equation}\label{HCform}
	K_T(\zeta)= e^{-\frac12 a^{\!\dag} \!a\, \kappa_o T}e^{a \zeta^*},
\end{equation}
conjugate to a \emph{Kraus-operator distribution function}, $D_T(\zeta)$, which evolves according to the equation
\begin{equation}\label{Autonomy}
	\boxed{
		\vphantom{\Bigg(}
		\hspace{15pt}
		\frac{\partial}{\partial t}D_t(\zeta) = \frac12\kappa(t)\!\left(\partial_{\zeta_1}^2+\partial_{\zeta_2}^2\right)\!D_t(\zeta),
		\hspace{10pt}
		\vphantom{\Bigg)}
	}
\end{equation}
where $\kappa(t) = \dot\Sigma(t)=\kappa_o e^{-\kappa_o t}$ is the effective observation rate and the initial condition is $D_0(\zeta)=\pi\delta^2(\zeta)$.
The exact same feature occurs here as in photodetection except that the Kraus operators redistribute according to a modified Brownian diffusion process (equation \ref{Autonomy}) rather than by a modified Poisson point process (equation \ref{Poisson}).
So this modification can be interpreted by exactly the same verse as before (modulo the change in instrument/stochastic process):
\begin{verse}\centering
	\fbox{\emph{The coupling of the heterodyne detector to the cavity is effectively screened over time.}}
\end{verse}
That is, the effective coupling at $t=T$ is equal to the coupling at $t=0$
rescaled by the renormalization factor, $Z_\alpha^{-1}=e^{-\kappa_o T}$.

This modified Brownian diffusion of the Kraus operators is a point that although obviously parallel to the main result of the Srinivas-Davies instrument appears not to be made at all in~\cite{wiseman1994quantum,Goetsch1994a,wiseman1996measurement}, although they obviously observe the relevance of Wiener-path integration.
On the other hand, Wiseman~\cite{wiseman1996measurement} does bring some attention to the Kraus-operator distribution function, specifically a generalization of $D_\infty(\zeta)$ considered for the purpose of adaptive measurement.
Indeed, it should be noted that~\cite{wiseman1996measurement} (unlike~\cite{wiseman1994quantum} and~\cite{Goetsch1994a}) is based on the realization that the registers of a measuring instrument, even in the adaptive setting, can be considered irrespective of the cavity state.
So it is fair to say that~\cite{wiseman1996measurement} is more-or-less aware of the principle of instrument autonomy, although it bypasses considering the evolution of the Kraus-operator distribution function.
Further, \cite{wiseman1996measurement} very clearly makes the point that the distribution of samples predicted by the Born rule can be considered to factor just as it does here,
\begin{equation}
\frac{d^2\zeta}{\pi}P_T(\zeta|\rho) = \frac{d^2\zeta}{\pi}D_T(\zeta)\Tr\left(K_T(\zeta)^{\!\dag} \!K_T(\zeta)\, \rho\right),
\end{equation}
where the Kraus-operator density $D_T(\zeta)$ is another example of the so-called method~C ostensible distribution function, as previously mentioned and as will be discussed in the Conclusion.

Unlike for photodetection, the ``screened diffusion process'' for heterodyne is actually similar to a more famous process, the Ornsetin-Uhlenbeck (OU) process~\cite{gardiner1986handbook}.
However, sample paths of the analogous OU process have solutions of the form,
\begin{equation}
	\nu\!\left[dw_{[0,T)}\right]=\int_0^T \!\!\sqrt{\kappa_o} dw_te^{-\frac12\kappa_o (T-t)},
\end{equation}
where the notable difference between these OU solutions and heterodyne, equation~\ref{heterofunc} rewritten here,
\begin{equation}
	\mu\!\left[dw_{[0,T)}\right]=\int_0^T \!\!\sqrt{\kappa_o} dw_te^{-\frac12\kappa_o t},
\end{equation}
is that the OU position essentially depends on the end of the record rather than the beginning.
Of course, that the heterodyne amplitude depends on just the beginning is important for the measurement interpretation as it implies that it is just the beginning of a record that reflects information about the state that stimulates it.
Finally, it can be noted that
\begin{align}
\frac{d^2\zeta}{\pi}D_T(\zeta)K_T(\zeta)^{\!\dag} \!K_T(\zeta)
&=\frac{d^2\zeta}{\pi}\frac{1}{\Sigma(T)}e^{-\zeta^*\zeta/\Sigma(T)}e^{a^\dagger\zeta}e^{- a^\dagger\! a\,\kappa_o T}e^{a\zeta^*}\label{firstformW}\\
&={\bf:}\frac{d^2\zeta}{\pi}\frac{1}{\Sigma(T)}e^{-\zeta^*\zeta/\Sigma(T)}e^{a^\dagger\zeta+a\zeta^*}e^{-a^\dagger\! a\,\Sigma(T)}{\bf:}\,.\label{secondformW}
\end{align}
Again, equation \ref{secondformW} makes obvious the POVM satisfied the completeness relation, and from equation \ref{firstformW} it is easy to see that the POVM elements are
\begin{equation}
D_T(\zeta)K_T(\zeta)^{\!\dag} \!K_T(\zeta)
=\proj{\zeta}
+O\left(|\zeta|^2e^{-\kappa_oT}\right),
\end{equation}
so that the time of Goetsch-Graham-Wiseman heterodyne instrument can be seen as a continuous bridge from the common idea of ``measurement in the coherent-state basis'' back to the equally important idea of not measuring.
In fact, the heterodyne instrument has an additional regularity to it that the Srinivas-Davies instrument does not, which is that the conjugate Kraus-operators are all close to the identity for small times.

\section{The heterodyne instrument alternative to energy quanta}\label{noquanta}

The Goetsch-Graham-Wiseman heterodyne instrument has many qualities to it that make it quite different from the traditional ideas of quantum measurement, and this section explains how the heterodyne instrument gets to the heart of quantum mechanics in several ways that are more operationally direct than the photodetector and the traditional idea of energy quantization.
There are three key features to point out:
The first feature is simply the limit of the heterodyne instrument, the informationally complete coherent-state POVM.
That the POVM is informationally complete means that this one instrument alone has the ability to see all the details of a quantum state (given enough copies of the state), unlike the photodetector which cannot see the phases of a number-state superposition.

The second feature of heterodyne is that it can be considered to evolve classically since the Kraus operators are infinitesimally generated.
In other words, equation~\ref{renormalizationW} can be thought of in terms of adjoint representation (or more famously the Baker-Campbell-Hausdorf theorem) and therefore understood to describe a (classical) differential geometric evolution in the shape of a specific 3-dimensional manifold.
That is, the motion infinitesimally generated by $a^\dag a$, $a$, and $-ia$ can be thought of as vector fields which flow along each other (i.e. pushforward) with derivatives Lie algebra-homomorphic to the (2+1)-dimensional affine group.
From this perspective, the Kraus operators march ballistically in the $a^\dag a$ direction while diffusing in the $a$ and $-ia$ directions with a diffusion rate that is shrinking because of the Lie derivative $[a^\dag a, a]=-a$.

The third feature is in the basic details of this differential-geometric motion, a detail that is in fact equivalent to the (obviously much more famous) quantization of the energy,
\begin{equation}\label{quantum}
	\Tr e^{-a^\dag\! a\, \kappa_o T} = \sum_{n=0}^\infty e^{-n\, \kappa_o T} = \frac{1}{1-e^{-\kappa_o T}}.
\end{equation}
Indeed, this will be the main point of discussion for this section.
Of course, one should expect that any logical interpretation and application of the canonical $[a,a^\dag]=1$ will have the same effect as energy quantization, so in an operator-algebraic sense there is no surprise here.
However, from the perspective of its physical interpretation, it is very interesting to see how the same effect as energy quantization appears when we are in fact not talking about energy or thermodynamic equilibrium at all.
That is, the reason the infinitesimal generator $a^\dag a$ actually appears in both the photodetector and heterodyne instrument is purely because of the complete positivity of the total operation.
Emphasized another way, $a^\dag a$ is not a Hamiltonian.

To be able to see this quantum feature alternatively as a differential geometric aspect of the heterodyne instrument will require an expression for the Kraus operators different from equation~\ref{HCform}, which, as it stands, does not respect the unitary left-invariance of the POVM.
For the purpose of respecting this left-invariant property, the Kraus operators of the heterodyne instrument can be transformed (in the co\"ordinate sense) by the equation
\begin{equation}\label{Cartan}
	\boxed{
		\vphantom{\Bigg(}
		\hspace{15pt}
		e^{-a^\dag \!a\, r}e^{a\zeta^*} = D_{\beta}e^{-a^\dag \!a\, r+1\zeta^*\zeta/2\Sigma_r}D_\alpha^\inv,
		\hspace{10pt}
		\vphantom{\Bigg)}
	}
\end{equation}
where defined are the unitary displacement operators,
\begin{equation}
	D_\alpha = e^{a^\dag \alpha - a \alpha^*},
\end{equation}
and two new complex amplitudes,
\begin{equation}\label{amplitudes}
	\alpha = \zeta/\Sigma_r
	\hspace{50pt}
	\text{and}
	\hspace{50pt}
	\beta = e^{-r}\alpha,
\end{equation}
with
\begin{equation}
	\Sigma_r = 1-e^{-2r}.
\end{equation}
%With expression \ref{Cartan}, the POVM elements take the symmetric form
%\begin{align}
%	e^{-\zeta^*\zeta/\Sigma(T)}e^{a^\dagger\zeta}e^{- a^\dagger\! a\,\kappa_o T}e^{a\zeta^*}
%	&=D_\alpha e^{-a^\dag \!a\, 2r+1\zeta^*\zeta/\Sigma_r}D_\alpha^\inv\\
%	&= e^{-(a-\alpha)^\dag \!(a-\alpha)\, 2r+1\alpha^*\alpha\Sigma_r}.
%\end{align}
The details of deriving equation~\ref{Cartan} will be left to Appendix~\ref{polar}.

Applying equation~\ref{Cartan} to the heterodyne instrument, where $r=\frac12\kappa_oT$,
it can be seen that the numerical factor cancels the exponential factor of the  Kraus-operator distribution (equation~\ref{Gaussian}) so that the heterodyne instrument transforms to
\begin{align}
	\frac{d^2\zeta}{\pi}D_T(\zeta)\;\O\Big(K_T(\zeta)\Big)
	&=\frac{d^2\zeta}{\pi}D_T(\zeta)e^{\zeta^*\zeta/\Sigma(T)}\;\O\Big(D_{\beta} e^{-a^\dag \!a\, r}D_\alpha^\inv\Big)\\
	&= \frac{d^2\zeta}{\pi}\frac{1}{\Sigma(T)}\;\O\Big(D_{\beta} e^{-a^\dag \!a\, r}D_\alpha^\inv\Big)\\
	&= \Sigma(T)\frac{d^2\alpha}{\pi}\;\O\Big(D_{\beta} e^{-a^\dag \!a\, r}D_\alpha^\inv\Big).
\end{align}
This means in turn that the POVM elements are
\begin{align}
\frac{d^2\zeta}{\pi}D_T(\zeta)K_T(\zeta)^\dag K_T(\zeta)
&= \Sigma(T)\frac{d^2\alpha}{\pi}\;D_\alpha e^{-a^\dag a\,\kappa_oT}D_\alpha^\inv,
\end{align}
which displays the unitary left-invariance,
and the completeness relation is thus
\begin{align}
	1
	&= \Sigma(T)\int\frac{d^2\alpha}{\pi}\;D_\alpha e^{-a^\dag a\,\kappa_oT}D_\alpha^\inv\\
	&= \Sigma(T)\int\frac{d^2\alpha}{\pi}\;D_\alpha^\inv e^{-a^\dag a\,\kappa_oT}D_\alpha.
\end{align}
Taking the groundstate (a.k.a. ``vacuum'' or ``highest weight state'') expectation of both sides and rearranging slightly gives
\begin{equation}\label{geometric}
	\int\frac{d^2\alpha}{\pi}\;\bra\alpha e^{-a^\dag a\,\kappa_oT}\ket\alpha
	= \frac{1}{\Sigma(T)},
\end{equation}
where defined are the famous coherent states,
\begin{equation}
	\ket\alpha \equiv D_\alpha \ket0,
\end{equation}
and $\ket0$ is the normalized groundstate.
Recognizing the left-hand-side of equation~\ref{geometric} as equal to the usual trace and substituting into the right-hand-side the expression for the covariance of the Kraus-operator distribution, we see that equation~\ref{geometric} is equivalent to equation~\ref{quantum}.
Once again, this means that the temporal behavior of the Kraus-operator distribution is in fact equivalent to energy quantization, except that the positive operator being traced is not a state and the parameter conjugate to $a^\dag a$ is not Plank's energy quantum multiplied by the inverse temperature.
Rather, the positive operator trace is from the completeness relation of the instrument and the parameter conjugate to $a^\dag a$ is the observation rate multiplied by the time.\\

Before concluding, it is worth paying some attention again to the Kraus-operator distribution with respect to the complex amplitudes of equation~\ref{amplitudes}, which I will rewrite here with the dependence in time,
\begin{equation}
	\alpha = \zeta/\Sigma(T)
	\hspace{50pt}
	\text{and}
	\hspace{50pt}
	\beta = e^{-\frac12\kappa_oT}\alpha,
\end{equation}
where again
\begin{equation}
	\Sigma(T) = \langle \zeta^*\zeta\rangle = 1-e^{-\kappa_o T}.
\end{equation}
Clearly, these have covariances
\begin{equation}
	\langle \alpha^*\alpha\rangle = \frac{\langle \zeta^*\zeta\rangle}{\Sigma(T)^2} = \frac{1}{\Sigma(T)}
	\hspace{50pt}
	\text{and}
	\hspace{50pt}
	\langle \beta^*\beta\rangle=e^{-\kappa_o T}\langle \alpha^*\alpha\rangle=\frac{1}{e^{\kappa_o T}-1}.
\end{equation}
So while the $\zeta$-covariance diffuses from the delta distribution to the normal distribution, the $\alpha$-covariance ``cools'' from the uniform distribution to the normal distribution and the $\beta$-covariance ``cools'' from the uniform distribution to the delta distribution.
Further, one can observe that the $\beta$-covariance cools in exact analogy to the Bose-Einstein occupation number.

\section{Conclusion: Trajectories, Measurement, and Instrument Autonomy}

In this article, I have discussed the theory of continual observation for two fundamental instruments: the Srinivas-Davies photodetector and the Goetsch-Graham-Wiseman heterodyne instrument.
The summary of both of these instruments centered around an idea I call the \emph{principle of instrument autonomy}, which focuses on the basic fact that a description of these theoretical instruments can be boiled down to an evolution equation for the Kraus-operator distribution function.
These evolutions shared a common modification to their respective Poisson and Wiener processes, a so-called \emph{screened observation rate} that comes from the \emph{effective amplitude renormalization} that the generators of both instruments feature.
It was then explained how the \emph{time} and \emph{positivity} of the continual heterodyne \emph{instrument} mirror completely the \emph{temperature} and \emph{Hamiltonian} of an equilibrium \emph{state} of light, demonstrating that the heterodyne instrument can be considered an alternative to energy quantization.

In particular, it must be noticed that sections~\ref{hetero} and~\ref{noquanta} managed to proceed without any reference to the eigenvalues of the operator $a^\dag a$.
In this sense, equations~\ref{renormalizationW}, \ref{Autonomy}, and~\ref{Cartan} were \emph{independent} of the Hilbert space --- i.e. autonomous.
Of course, these eigenvalues are already uniquely defined by the commutator $[a^\dag a, a]=-a$ and existence of the groundstate $a\ket{0}=0$, yet still the groundstate was not required until the very end.
Stepping away from Hilbert space has no obvious advantage for the Weyl-Heisenberg observables $Q$ and $P$ since the Hilbert space is essentially uniquely determined (as is more famously asserted by the Stone-von Neumann theorem.)
Yet this Hilbert-space-independent or autonomous style of considering the heterodyne instrument still has the real benefit of offering to the mind operational meanings of the complex operator algebraic structure that are otherwise almost purely associative.
Meanwhile, this ability to remove the instrument from Hilbert space is actually far more revolutionary in the context of measuring spin systems because the quantization of spin is not uniquely defined by the Lie algebra of Spin observables~\cite{Jackson2021perform}.\\

Also important to point out is that these instrument autonomous methods have been touched on by Wiseman~\cite{wiseman1996measurement} with the discussion of methods A, B, and C of choosing an ostensible distribution for simulating measurement.
There, what was recognized is that measurement records do not have to be sampled from the Born rule, the so-called method A.
The very opposite of this, called method B, was to consider measurement records sampled from a uniform distribution.
While method A is obviously state-dependent, method B is obviously state-independent.
Method C was then generally defined to be anything in between these two methods.
However, the method C actually used for considering heterodyne, the so-called ``half-way house'', is not so in between after all in the sense that it still results in sampling another state-independent distribution function, a variant of the Kraus-operator distribution function discussed in this article.
What that discussion illustrates is that state-independent ostensible distributions do not in themselves demonstrate instrument autonomous because the distributions of method B are both hard to sample and also have no sense of the structure of the instrument.
However, Wiseman's method C ``half-way house'' \emph{does} demonstrate instrument autonomy as this current article has shown.\\

I believe the understanding of heterodyne instruments illustrated in this article shows that the principle of instrument autonomy and the Kraus-operator distribution function evolution have an enormous potential for understanding the role of instruments in the foundations of quantum mechanics, and I would like to mention three places where this principle and the Kraus-operator evolution can be further applied:
First of all, this idea of autonomous evolution has already demonstrated the ability to explain how a ``measurement in the spin-coherent-state basis'' could be realized~\cite{Jackson2021perform}.
Second, the time of these autonomous evolutions strongly resembles the famous ``time'' dimension of modern quantum condensed matter theory~\cite{matsubara1955new,klauder1960action,altland2010condensed} and therefore could offer a more operational interpretation of its meaning, perhaps more satisfying than purely associative algebra.
Third, the time of these autonomous evolutions (especially for informationally complete instruments) could allow one to define principle-based connections along the world lines of general relativity, perhaps offering insights into quantum gravity.

\acknowledgements{
	The author would like to thank Carl Caves and Mohan Sarovar for all of the helpful discussions and financial support.
	This work was supported by the U.S. Department of Energy, Office of Science, Office of Advanced Scientific Computing Research, under the Quantum Computing Application Teams (QCAT) program. Sandia National Laboratories is a multimission laboratory managed and operated by NTESS, LLC., a wholly owned subsidiary of Honeywell International, Inc., for the U.S. DOE's NNSA under contract DE-NA-0003525. This paper describes objective technical results and analysis. Any subjective views or opinions that might be expressed in the paper do not necessarily represent the views of the U.S. Department of Energy or the United States Government.}

\bibliography{PhotoAutonomy}

\appendix

\section{Deriving the coordinate transformation of the heterodyne operations}\label{polar}

In this appendix, equation~\ref{Cartan} is derived.
It should be noticed that the right-hand-side of equation~\ref{Cartan} can be recognized as a singular-value decomposition.
However, literally considering singular values is not very useful for establishing equation~\ref{Cartan}; far more useful is to consider the differential geometric idea of the Lie derivative.
In this way, the right-hand-side of equation~\ref{Cartan} is more similar to what is known as a Cartan decomposition, in which case the POVM is similar to a (non-Riemannian) symmetric space.
As such, the singular-value decomposition can be understood as a matrix representation of the more fundamental Cartan decomposition.
Further discussion of the Cartan decomposition can be found in~\cite{Jackson2021perform}.
Depth aside, it can simply be appreciated that the method here is not based on singular values but rather on splitting exponentials and commuting them through each other.

Define the unitary displacement operators
\begin{equation}
	D_\alpha = e^{a^\dag \alpha - a \alpha^*}.
\end{equation}
We will show that
\begin{equation}
	D_\beta e^{-a^\dag \!a\, r}D_\alpha^\inv = e^{-\frac 12(V+i\phi)}e^{a^\dag\nu}e^{-a^\dag \!a\, r}e^{a\mu^*}
\end{equation}
by finding expressions for the parameters on the right.
Observe
\begin{align}
	D_\beta e^{-a^\dag \!a\, r}D_\alpha^\inv
	&=e^{-\frac12\beta^*\beta-\frac12\alpha^*\alpha}(e^{a^\dag\beta}e^{-a\beta^*})e^{-a^\dag \!a\, r}(e^{-a^\dag\alpha}e^{a\alpha^*})\\
	&=e^{-\frac12\beta^*\beta-\frac12\alpha^*\alpha}e^{a^\dag\beta}e^{-a\beta^*}(e^{-a^\dag e^{-r}\alpha}e^{-a^\dag \!a\, r})e^{a\alpha^*}\\
	&=e^{-\frac12\beta^*\beta-\frac12\alpha^*\alpha+e^{-r}\beta^*\alpha}e^{a^\dag\beta}(e^{-a^\dag e^{-r}\alpha}e^{-a\beta^*})e^{-a^\dag \!a\, r}e^{a\alpha^*}\\
	&=e^{-\frac12\beta^*\beta-\frac12\alpha^*\alpha+e^{-r}\beta^*\alpha}e^{a^\dag(\beta- e^{-r}\alpha)}e^{-a^\dag \!a\, r}e^{a(\alpha-e^{-r}\beta)^*}.
\end{align}
Reading the arguments of these exponentials gives
\begin{equation}
	\nu = \beta-e^{-r}\alpha
	\hspace{50pt}
	\text{,}
	\hspace{50pt}
	\mu = \alpha - e^{-r}\beta
	\hspace{50pt}
	\text{,}
\end{equation}
\begin{equation}
	V= \beta^*\beta+\alpha^*\alpha-e^{-r}(\beta^*\alpha+\alpha^*\beta)
	\hspace{50pt}
	\text{, and}
	\hspace{50pt}
	i\phi = e^{-r}(\beta^*\alpha-\alpha^*\beta).
\end{equation}
In particular, we are interested in the operator $e^{-H_o r}e^{a\zeta^*}$ where $\nu=0$, and therefore these equations become
\begin{equation}
	\beta=e^{-r}\alpha\,
	\hspace{50pt}
	\text{,}
	\hspace{50pt}
	\zeta = (1 - e^{-2r})\alpha
	\hspace{50pt}
	\text{,}
\end{equation}
\begin{equation}
	V= (1-e^{-2r})\alpha^*\alpha
	\hspace{50pt}
	\text{, and}
	\hspace{50pt}
	i\phi = 0.
\end{equation}
Rearranging, we see the desired expression
\begin{equation}
	\boxed{
		\vphantom{\Bigg(}
		\hspace{15pt}
		e^{-a^\dag \!a\, r}e^{a\zeta^*} = D_{\beta}e^{-a^\dag \!a\, r+1\zeta^*\zeta/2\Sigma_r}D_\alpha^\inv,
		\hspace{10pt}
		\vphantom{\Bigg)}
	}
\end{equation}
where
\begin{equation}
	\alpha = \zeta/\Sigma_r
	\hspace{50pt}
	\text{and}
	\hspace{50pt}
	\beta = e^{-r}\alpha,
\end{equation}
with
\begin{equation}
	\Sigma_r = 1-e^{-2r}.
\end{equation}

\end{document}